\documentclass{article}
\usepackage{ijcai23}

\usepackage{times}
\usepackage{soul}
\usepackage{url}
\usepackage[hidelinks]{hyperref}
\usepackage[utf8]{inputenc}
\usepackage[small]{caption}
\usepackage{graphicx}
\usepackage{amsthm}
\usepackage{amsmath, amssymb}
\usepackage{booktabs}
\usepackage{algorithm}
\usepackage{algorithmic}
\usepackage[switch]{lineno}
\usepackage{subfigure}
\usepackage{comment}
\usepackage{float}
\newcommand{\quotes}[1]{``#1''}

\title{Integration of Deep Reinforcement Learning and Agent-based Simulation to Explore Strategies Counteracting Information Disorder}

\author{
Luigi Lomasto$^1$\and
Andrea Camoia$^1$\and
Alfonso Guarino$^1$\and
Nicola Lettieri$^2$\and
Delfina Malandrino$^1$\And
Rocco Zaccagnino$^1$\and
\affiliations
$^1$Department of Computer Science, University of Salerno, Via Giovanni Paolo II, 130, Fisciano, SA, Italy\\
$^2$National Institute for Public Policy Analysis (INAPP), Rome, Italy
\emails
llomasto@unisa.it,
a.camoia@studenti.unisa.it,
aguarini@unisa.it,
n.lettieri@inapp.org, 
dmalandrino@unisa.it,
rzaccagnino@unisa.it,
}

\begin{document}

\maketitle

\begin{abstract}
In recent years, the spread of \textit{fake news} has triggered a growing interest in Information Disorders (ID) on social media, a phenomenon that has become a focal point of research across fields ranging from complexity theory and computer science to cognitive sciences. Overall, such a body of research can be traced back to two main approaches. On the one hand, there are works focused on exploiting data mining to analyze the content of news and related metadata (\textit{data-driven} approach); on the other hand, works are aiming at making sense of the phenomenon at hand and their evolution using explicit simulation models (\textit{model-driven} approach). In this paper, we integrate these approaches to explore strategies for counteracting IDs. Heading in this direction, we put together: \textit{i.} an Agent-Based model to simulate in a scientifically sound way both complex fake news dynamics and the effects produced by containment strategies therein; \textit{ii.} \textit{Deep Reinforcement Learning} to learn the strategies that can better mitigate the spread of misinformation. The outcomes of our work unfold on different levels. From a substantive point of view, the results of preliminary experiments started providing interesting cues about the conditions under which given policies can mitigate the spread of misinformation. From a technical and methodological point of view, we scratched the surface of promising and worthy research topics like the integration of social simulation and artificial intelligence and the enhancement of social science simulation environments.
\end{abstract}

\noindent\textbf{\textit{This paper is a preliminary version of the published paper \cite{zaccagnino2025turning}. \textit{Turning AI into a regulatory sandbox: exploring information disorder mitigation strategies with ABM and deep reinforcement learning}, Neural Computing and Applications. DOI: \href{https://doi.org/10.1007/s00521-025-11342-y}{https://doi.org/10.1007/s00521-025-11342-y}. Please cite the published version.}}

\section{Introduction}
\label{sec:introduction}
In the era of digitization, social networks have become powerful platforms for the dissemination of information, connecting individuals across the globe in unprecedented ways. However, alongside the benefits, these platforms have been confronted with a pervasive challenge: Information Disorder (ID). This phenomenon, whose interest has been triggered by the spread of "fake news," encompasses also misinformation and disinformation, collectively contributing to an environment often characterized by confusion, deception, and discord. ID on social networks not only undermines the credibility of digital media but also poses significant threats to public discourse, democratic processes, and societal harmony as explained in \cite{allcott2017social}. In addition, the psychological aspect is also crucial to understanding user behavior as in \cite{pennycook2021psychology}. 

The ease with which misleading content can be propagated and amplified has urged the exploration of research approaches allowing to understand the phenomenon at hand and its dynamics before trying to mitigate and counteract it.

In this respect, it is worth considering the most recent developments in the field of computational social sciences where two main approaches to the study of social phenomena can be identified: a \textit{data-driven} approach and a \textit{model-driven} one \cite{conte2014agent}. In the former, huge amounts of data are analyzed to identify correlations and patterns, attempting to deduce the causes of social phenomena based on observed evidence. In the latter, simulation models, in particular agent-based ones (ABMs), are used to create generative explanations of social phenomena. Such models incorporate the description of both internal and external mechanisms supposed to drive the phenomena, offering a deeper understanding of the underlying dynamics. While the \textit{data-driven} approach relies on correlations in existing data, the \textit{model-driven} one focuses on creating theoretical models that better allow exploration, understanding, and prediction of complex social phenomena and dynamics of behavioral change.

In such a scenario, the integration of the \textit{data-driven} and \textit{model-driven} approach appears to be a promising strategy. On the one hand, simulations can provide a realistic and dynamic context for exploring complex social behaviors and responses to them. On the other hand, machine learning can be used for a variety of purposes: to analyze data generated by simulations and identify hidden patterns and trends in them; to assess the effectiveness of responses to target phenomena in simulated scenarios; to endow agents with a higher level of "intelligence" or cognitive complexity.

All of the above holds true for fake news and information disorders as it emerges from an even cursory review.

\textit{Data-driven} approaches consist of AI solutions usually based on machine learning (ML), deep learning (DL), and natural language processing models. The common idea is to detect fake news using text classifier models trained to distinguish fake content on social networks. In~\cite{capuano2023content}, ML and DL algorithms like Support Vector Classifier, Random Forest, Logistic Regression, LSTM, RNN, and pre-trained models like BERT, RoBERTa, DistilBERT ELMo, etc. are listed as primary and standard solutions to perform embeddings and classification activities for fake and real news. Unfortunately, solutions based only on news content 
may be ineffective for breaking news. 
To overcome this limitation, several innovative solutions have been proposed.  As an example, in~\cite{mosallanezhad2022domain}, a Reinforcement Learning method is used to modify the representation of news to classify fake content by moving from one domain to another. In other cases, user sentiment analysis~\cite{alonso2021sentiment} and fine-tuned BERT models~\cite{alawadh2023attention} are used as main features to improve and outperform standard ML classifiers. Finally, novel approaches that consider labeled and unlabeled data are proposed. In~\cite{Dong20234758}, the authors propose a framework that combines Deep Neural Networks models trained to learn for implicit relationships between labeled data, with clustering techniques used to learn for explicit relationships between unlabeled data. Despite such considerable efforts, the main limitation is related to the lack of a knowledge base for newly published news, which is the first problem for these types of solutions. 

As for the \textit{model-driven} approaches, research studies on online information disorder focus on the concepts of \textit{Echo Chamber} and \textit{Filter Bubble}~\cite{morini2021toward,flaxman2016filter}, which are generally used to describe the isolation of users on social networks and to study the spread of misinformation therein. In~\cite{abbruzzese2021detecting,gaeta2023novel}, the approximate rough set theory has been used to model and reason about how user groups share information and how a set of news circulates in social networks. An obvious limitation of these solutions is the inability to determine why a news item does or does not influence user opinion, since the content of the news item is not evaluated.  
In~\cite{malecki2022multi}, a multi-agent-based model is used to study the spreading of fake information on Scale-Free Networks. The authors show how increasing agent reputation encourages the spread of false information.
In~\cite{alassad2023developing}, the authors introduced a study that presents a systematic and multidisciplinary agent-based model aimed at interpreting and simplifying the dynamic actions of users and communities within evolving online social networks. The goal of this approach is to control and monitor the spread of malicious information between communities by using an Organizational Cybernetic Approach. The results are promising if we consider the analysis of information spread but have limitations related to users' beliefs. 
Similar research~\cite{gausen2021can}, based on the agent model, aims to define countermeasures and understand their effects on the spread of fake news.
Recently, research activities focused on balanced influence maximization on social networks have been performed. In fact, in~\cite{yang2023balanced} the researchers propose a Balanced Influence Maximization framework based on Deep Reinforcement Learning composed of two modules, the first one to evaluate the impact of entity correlation on information propagation considering the users' behavior, and the second one based on a deep reinforcement learning (DRL) that aim to found seed nodes useful to maximize the balanced influence. 
These latest trends, based on agent-based models (ABM) or DRL, represent a novel approach to study and develop solutions to understand and counter information disorder. 

Despite the number of works published on the subject matter, as far as we know, there are few attempts to integrate inductive inferences provided by \textit{data-driven} approaches with the insights coming from the explicit operational description of the target phenomenon that characterizes simulation models.
Considering different scenarios, in the last years hybrid approaches to solve different problems are proposed as \cite{lettieri2023knowledge} and \cite{cozza2020hybrid}. 
In light of these considerations, this paper introduces a novel layered framework designed to delve into the intricacies of ID within social networks. The objective is to grasp its propagation mechanisms and explore potential remedies. The proposed framework unfolds on two distinct tiers: 
\begin{itemize}
\item \textit{model-driven tier} (MDT): includes the design and implementation of a particular model to replicate the evolution of the phenomenon, generating \textit{simulation data content}, which refers to the information that outlines the specifics of the simulation.
\item \textit{data-driven tier} (DDT): higher tier that uses a deep reinforcement learning-based agent that exploits the content of the simulation data produced by the MDT to identify optimal actions aimed at mitigating the impact of the ID.
\end{itemize}

The outcomes from initial experiments conducted to evaluate the efficacy of the proposed framework reveal intriguing insights into the circumstances under which specific policies can mitigate the spread of misinformation. This underscores the importance of interdisciplinary collaboration to combat information disorder and sets the stage for future research and technological advancements in this crucial field.

The remaining part of the paper is organized as follows: Section~\ref{sec:Framework} offers details on our approach, Section~\ref{sec:experiments} presents the experiments and the results so far achieved, and, lastly, Section~\ref{sec:conclusion} is devoted to final remarks. 

\section{ID phenomenon: model definition}
\label{sec:problem}

To describe the ID phenomenon, we started from the model proposed in~\cite{tornberg2018echo}, where the authors \quotes{shaped} the simulation considering the \textit{echo chambers} phenomenon. An echo chamber is defined as a set of users characterized by two important properties: \textit{opinion polarization} $P_N$ and \textit{network polarization} $P_0$. Opinion polarization provides a measure of how much users share content on a specific topic with similar points of view, whereas network polarization refers to the extent to which individuals or groups within a network are divided into opposing views or ideologies (e.g., high density of connection between users inside the echo chamber and little connections with users outside). 
To emulate the echo chamber phenomenon in a network, the following parameters were defined: the average degree of links for the nodes $k$, the activation threshold for information diffusion $\theta$, the number of nodes that belong to the echo chamber $c$, the total number of edges $E$, and the total number of nodes $N$. By using a combination of such parameters and defined policies, the authors run the model of fake news propagation starting from an Erdős-Rényi network structure~\cite{erdds1959random}. At the end of the procedure, the \textit{virality} function $V=(\theta,P_n,P_O)$ is defined to describe the percentage of nodes activated so as to study the correlation between $V$, $P_N$, and $P_O$. If the virality goes above the critical threshold of $0.5$ the fake news has spread across the majority of nodes.

Inspired by the aforementioned model, we have modeled the problem defining two types of agents: \textit{basic agent} and \textit{Super-Agent} described below. The cooperation between basic agents and the Super-Agent gives rise to a framework based on model-driven and data-driven tiers. 

The proposed model assumes the presence of an echo chamber within a network that starts the spread of fake news. Basic agents (nodes) in the network can have three different types of opinion: \textit{A} (supporting \textit{fake} news), \textit{B} (supporting \textit{true} news), and \textit{neutral} (no opinion). Each node is characterized by a \textit{activation threshold} $\theta$, which informally represents the \quotes{credulity}, and plays a crucial role in opinion dynamics. Network equilibrium depends on basic agents, which represent the real users of an SN, that try to convince their neighbors to adopt their supported opinion, while the Super-Agent, which for example could be a governance agency, can shift these balances by persuading numerous basic agents to support opinion \textit{B}. The Super-Agent is the only one that identifies and intervenes with nodes disseminating type \textit{A} news. In brief, the basic agents aim to spread their opinion, while the Super-Agent tries to reduce the fake news propagation (measured as a \textit{global cascade}, whose details will be given in the following Sections) by encouraging basic agents to adopt opinion \textit{B}. The Super-Agent learns from the simulation data utilizing a DRL technique to predict future actions and detect network changes. The initialization of the echo chamber involves selecting a random node as the starting point for opinion \textit{A}, making all connected nodes support opinion \textit{A}. Similarly, outside the echo chamber, a random node is selected as the starting point for opinion \textit{B}, activating all connected nodes to support opinion \textit{B}. The synergy among fundamental agents and the Super-Agent encapsulates the ID phenomenon, constituting the basic tier within a conceptual infrastructure. This infrastructure includes the \textit{Model-Driven Tier} (MDT) and the \textit{Data-Driven Tier} (DDT), both orchestrated to achieve the dual objectives of offering insights into the ID and proposing potential remedial measures. 

\section{Framework Description}
\label{sec:Framework}
\begin{figure}[h!t]
    \centering
    \includegraphics[width=8cm]{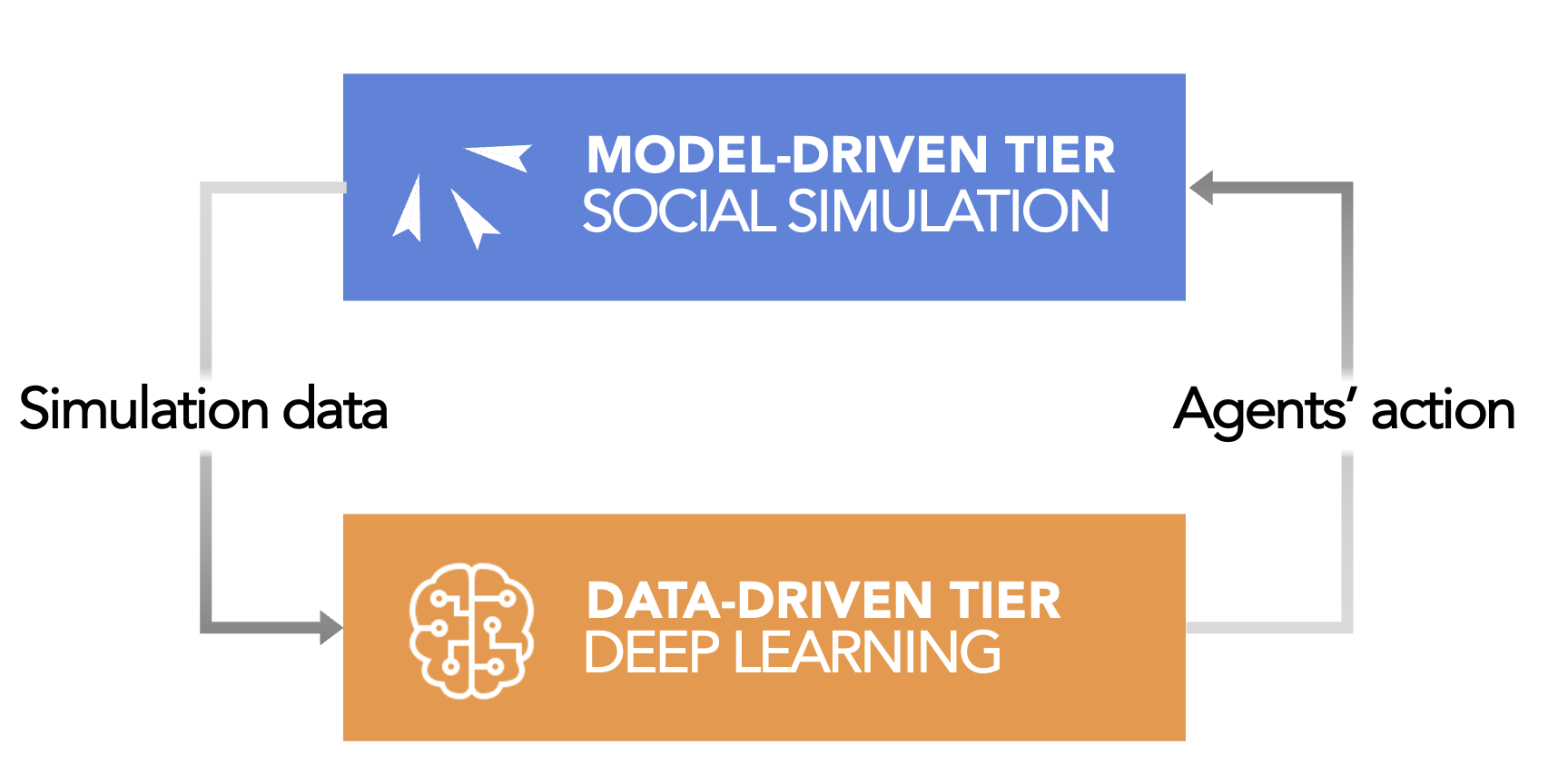}
    \caption{Visual abstract of tiers'architecture}
    \label{fig:visualabstract}
\end{figure}

In this section, we will provide details about the proposed layered framework  (shown in Figure \ref{fig:visualabstract}), outlining the objectives of each tier, and how each tier interacts with the tier above it.
\subsection{Model-Driven Tier (MDT)}
\label{ssec:abmcomp}
The MDT comprises an ABM constructed based on the phenomenon designed in Section \ref{sec:problem}. During the simulation process, MDT provides 
information pertinent to network dynamics (\textit{simulation data content}) and news propagation to the DDT (see Section \ref{ssec:drlcomp}).   As described above, the basic agents, that depict the users on social networks, can have three types of opinions: opinion \textit{A} (supporting fake news), opinion \textit{B} (supporting real news), and opinion \textit{C}  (no opinion). 

To start the simulation, first, a Erdős-Rényi network is created randomly as proposed in~\cite{tornberg2018echo}.

In summary, first, an echo chamber is created as follows.
The initial values (in $[0,1]$) for $P_O$, $P_N$, and for \textit{ECF}, i.e., the \textit{echo chamber fraction} parameter which indicates the fraction of nodes that should participate in the echo chamber, are set. Then, the number of nodes to be included in the echo chamber, denoted as $ c $, is calculated as $c = N \times \text{ECF}$. Thus, the $ c $ nodes forming the echo chamber are randomly selected from the network and their attribute \texttt{is-in-cluster} is set to true.
To complete the network construction, the echo chamber nodes are connected to the remaining nodes as follows.
To this aim, the number of edges to be used for connecting the echo chamber with the outside nodes, denoted with $ E' $, is given by $E' = E \times P_N$, where $E$ is the number of edges of the network.
To increase the degree of centrality among nodes within the echo chamber and promote cohesion, a random check is performed on $ E' $. 
More in detail, if for a random edge in $E'$, only one of the two connected nodes belongs to the echo chamber (call this node T for simplicity), the edge is deleted and replaced with another edge to connect T with another node that belongs to the echo chamber. 
Next, a random node \(init\) within the echo chamber is selected as the activation point, and its \texttt{is-a-active} attribute is set to true. A random value for the Opinion Metric (OM) is assigned to \(init\). Then, the neighbors of \(init\) are also selected, and their \texttt{is-a-active} attributes are set to true.
We proceed in a mirrored way for nodes of type B. 
Finally, $\Theta$ is set for each node in the network. Nodes belonging to the echo chamber have their $\Theta $ adjusted to $\Theta - P_O $, while nodes outside simply retain the initial $\Theta $ value chosen at the start of the simulation. This means that nodes within the cluster are more influenced by opinions.

\subsection{Data-Driven Tier (DDT)}
\label{ssec:drlcomp}
To counter the spread of fake news, we have defined a Super-Agent based on deep Q-learning~\cite{clifton2020q}, a DRL algorithm, connected to the agent-based model tier. Considering the information coming to MDT, it aims to understand the state of the network and take the best action for mitigating the spread of A opinions. 
We opted for a DRL approach because, through the Q-learning algorithm, it would become quite tough to map actions and rewards for a set of so large states. Moreover, in a preliminary phase of the project, we observed that such a model overcame policy-based approaches. 
The Super-Agent is responsible for intervening in the simulation to mitigate the spread of fake news. In this section, we explain the parameters used for the DRL tier and the possible actions that the Super-Agent can perform. It works as a supervisor controlling the network changes related to the spread of fake news introduced during the simulation.  
The parameters observed by the Super-Agent during the simulation are \textit{Global\_Cascade}, \textit{Global\_Opinion\_Metric} and  \textit{Most\_Influent\_A\_nodes} and are described below. 

 \textbf{\textit{Global Opinion Metric (GOM)}}: it indicates the global average of \textit{OM} values for each \textit{basic-agent} in the simulation. It is computed as 
\begin{equation}
            GOM = \frac{\sum_{1}^{N} OM(ba_i) }{N}
\end{equation}
where the \textit{Opinion Metric (OM)} is defined as follow. Each basic-agent (ba) has an OM, used to indicate the idea that they have, which can take the following values: 
\begin{equation}
OM = \begin{cases}
0 \le x \le 0.33 & \text{Opinion B}\\
0.33 < x < 0.66 & \text{Opinion C}\\
0.66 \le x \le 1 & \text{Opinion A}\\
\end{cases}
\end{equation} 
\textbf{\textit{Global Cascade (GC)}:} denotes the fraction of basic agent (ba) of type A actives. It is computed as: 

\begin{equation}
GC = \frac{\sum_{1}^{N} is-a-active(ba_i)}{N}
\end{equation}

 \textbf{\textit{Most Influent A nodes (MIA):}}  The fraction of the most influential nodes of type A, ordered in descending order of \texttt{betweenness centrality} (or \texttt{degree centrality}) \cite{freeman1978centrality};

 Other parameters used are \textit{node-range} which indicates the fraction of nodes considered by the Super-Agent for performing specific actions, \textit{node-range-static-b} indicating the fraction of nodes to which the Super-Agent forces the maintenance of opinion B, \textit{choose-method} which determines the method of choice of links that the Super-Agent creates, based on degree, betweenness, etc., and \textit{global-warning} indicating whether the warning can be global or not.
The actions the Super-Agent can perform are: \textit{Warning}, \textit{Reiterating}, \textit{Forcing} (all inspired by~\cite{ruffo2023studying}), and \textit{Observing}. 

 \textbf{Warning Action}. It can be either global or localized, impacting the basic agents differently:

\begin{itemize}
    \item \textit{Global Warning}: When activated, this type of Warning sets the warning variable to true for all basic agents across the network. Once triggered, this global warning state persists until the end of the simulation. The logic of this approach is to uniformly affect all agents in the network, independently of their individual properties or states.
    \item \textit{Local Warning}: Contrarily to the \textit{Global Warning}, the localized version targets only a select group of basic agents. This selective approach allows for a more nuanced and targeted influence on the network, impacting only those agents meeting certain criteria.
\end{itemize}
Once the warning variable is activated on a basic agent, it will be active for the whole simulation. 
This action works as it follows: when a basic agent is about to be influenced by an opinion, it checks the state of its warning variable. If the value is true, the model generates a random number \textit{x} in the range $\mathbb{R}[0,1]$. This number is then compared to the predefined warning-impact threshold set at the beginning of the simulation. If the number generated is less than or equal to the threshold, the model starts the opinion update procedure.
Through this alert mechanism, the simulation effectively models the dynamics of opinion formation and change, considering both global influences and more localized impacts. Concerning the real world, the warning action can be seen as a system warning that the news does not have reliable evidence. 

\textbf{Reiterating Action}. The reiterate action sets to \textit{True} the reiterate variable for a set of chosen basic agents. When a basic agent is about to be affected by opinion, it checks the reiterate variable. The basic-agents with reiterate variable set to \textit{True} receive news of type B during the next ticks of the simulation for some times equal to the number of neighbors. During this process, a random number \textit{y} in range $\mathbb{R}[0,1]$ is generated. If the \textit{y} value is less than $\theta$, the \textit{OM} is updated to become more close to B opinion. If the basic-agent changes its opinion, the reiterate process can stop before the established ticks. In this case, considering the real world, real news is shown several times to encourage a change of opinion. 

\textbf{Forcing Action}. This action is used to force specific basic agents to maintain the B opinion during the entire simulation process. The choice of the basic agents is achieved through the application of one of three distinct selection criteria, which are \textit{betweenness, page rank, and degree centrality}, selecting nodes with the highest values of such measures.
The number of basic agents is determined by the \textit{node-range-static-b} parameter, which represents a fraction of the total nodes available in the system. For instance, if \textit{node-range-static-b} is set to 0.1, it implies that 10\% of the total nodes will be selected.
The essence of this action lies in altering the \textit{opinion-metric} of the selected nodes, setting it to 0 and setting \textit{is-opinion-b-static} variable to True for the selected basic agents. Consequently, these nodes are effectively \textit{locked} into maintaining the B opinion, regardless of any incoming information of type A that they may encounter during the simulation. In the real world, generally, there are trusted actors, like online newspapers, who can make a relevant contribution to the spread of real news. 

\textbf{Observing Action}. In this case, the Super-Agent observes the phenomenon without acting. Generally used when the network conditions are acceptable. 

A reward function \textbf{R}  for the learning process has to be defined. In our case, R is performed by considering the three parameters \textit{GC}, \textit{GOM}, and \textit{MIA}, starting from the first \textit{tick} of the simulation. The rewards are the result of a series of empirical evaluations in a preliminary observation phase. The \textit{action-weight (AW)} value is computed as $AW = \frac{1}{GOM} + \frac{1}{MIA} $. The inverse of \textit{GOM} and \textit{MIA} are considered because a low value (close to zero) for both, indicates a high spread of real news and a low spread of fake news.  
Following, the \textit{Action-Result (AR)} is computed as the difference of the \textit{GC} values between the current tick (k) and the previous one (k-1): $AR = GC_{k} - GC_{k-1}$. This parameter is used to understand if the number of agents supporting opinion A has increased during the simulation.
The AR parameter can assume three different meanings related to its value: 
\begin{equation}
\begin{cases}
AR < 0 & \text{(the GC improves)}\\
AR = 0 & \text{(the GC hasn't changed)}\\
AR > 0 & \text{(the GC worsens)}\\
\end{cases}
\end{equation} 
Considering the parameters just described, a reward for the model is computed as follows. If the Super-Agent has acted with a positive or neutral impact on $GC$ ($AR \le 0$), the reward is calculated as $R = (1+AW \times 0.5)-AR$.
In contrast, if the $GC$ worsens ($AR>0$) the reward becomes: $R = (0+AW \times 0.5)-AR$.
Obviously, in cases where the Super-Agent has already executed the \textit{Warning} or \textit{Forcing} actions and executes them other times, the reward value is evaluated differently. Specifically, for these cases if $\textit{GC} > 0.5$ then $R=0$, $R=1$ otherwise. 

 \textbf{Architecture.}  The model's architecture includes one dense layer of size 24 with \textit{relu} activation, one dense layer of size 12 with \textit{relu} activation, and a last dense layer of size 4 (i.e., the number of actions) with \textit{linear} activation. The batch size is put at 128.

\subsection{Practical Aspects}
\label{ssec:practical}
In this section, we sketch the practical aspects of our approach which is based on the combined use of \textit{Python} and \textit{NetLogo}, exploiting the NetLogo API for Python, i.e. \textit{PyNetLogo}~\cite{jaxa-rozen2018}. 
At the beginning, the initialization of the simulation parameters occurs through a \textit{Python} script that sends the settings to \textit{NetLogo} and starts the simulation. Subsequently, once the simulation has started, observations on the network and its agents are collected from the \textit{NetLogo} environment; at each \textit{tick} the relevant parameters of the simulated environment are taken and exploited by the \textit{Super-Agent} to take decisions that impact the simulation and its agents. From these operations, the \textit{Super-Agent} obtains a reward that will better guide future decisions (in future \textit{tick}).
At the end of the simulation, all the parameters of interest to the researcher are taken and made accessible through intuitive infographics.
This combined strategy will be used both to carry out experiments involving the \textit{Super-Agent} (in the training phase as well as in the testing phase) and for all those tests that do not involve this entity.

\subsection{Network Initialization and Simulation concepts}
\label{ssec:netinit}
The network initialization phase starts by setting the following parameters: \textit{(i)} $N$ the number of nodes (basic agents); \textit{(ii)} the network topology (i.e., Erdős-Rényi network, Small World or Preferential Attachment); \textit{(iii)} the number of ticks; \textit{(iv)} $\theta, P_N, P_O $, as in~\cite{tornberg2018echo}    \textit{(v)} echo-chamber-fraction: the fraction of nodes that belong to the echo chamber; \textit{(vi)} initial-opinion-metric: the initial value of OM for neutral agents; opinion-metric-step (OMS): the step value for updating the OM; \textit{(vii)} k, is-a-active. During the simulation, other parameters are set and/or changed like \textit{Warning Impact} used to define how much the warning action impacts the nodes, 
\textit{node-range} to indicate the fraction of nodes that are linked with the Super-Agent, \textit{node-range-static-b} used to define the fraction of nodes to which the Super-Agent forces to maintain the opinion B,  \textit{choose-method} that determines the links created by Super-Agent considering \textit{degree, betweenness, and page-rank} values and \textit{global-warning} to indicate if the warning is global for the network or not and \textit{sa-delay}, to define the action frequency of the Super-Agent.  
Throughout the simulation, at every tick, each basic agent checks its neighbors, and if the fraction of neighbors supporting a specific opinion is higher than its own $\theta$ it changes its own opinion accordingly.
As explained in Section~\ref{ssec:drlcomp}, during the simulation, the Super-Agent can perform three important actions: \textit{Warning},  \textit{Reiterate} and \textit{Forcing} or may decide not to take action.

\section{Experiments and Results}
\label{sec:experiments}
In this section, we show experiments and results obtained from the simulations with and without the Super-Agent. The goal is to show how the super-agent actions are useful to counteracting and limiting the fake news spread phenomenon, under specific conditions. The parameters used to evaluate, understand, and explain the problem are $P_N$, $P_O$, $\theta$ previously seen, and Virality ($V$), defined below. Credulity is high if $\theta$ is low and vice versa. $V$ is defined as the fraction between the number of times the GC is greater than $0.5$ and the total of simulations \textit{T} runs.
The Virality function \textbf{V} is defined as:
\begin{equation}
 V = \frac{\sum_{1}^{T} CGC(global\_cascade_i)}{T} 
\end{equation}
\begin{equation}
CGC(global\_cascade_i) = 
\begin{cases}
GC_i > 0.5 & \text{return 1} \\
GC_i \le 0.5 & \text{return 0} \\
\end{cases}
\end{equation}

The objective of the experiments is to compare V values obtained from the simulations (as in~\cite{tornberg2018echo}) with and without the Super-agent,  changing $P_N$, $P_O$, and  $\theta$, to comprehend if the Super-Agent is effective or not and what is its impact in countering fake news propagation.
We remark that the critical value for V is 0.5 (as indicated by~\cite{tornberg2018echo}).

\begin{table}[H]
    \centering
    \begin{tabular}{lrr}
        \toprule
        \textbf{Attribute}  & \textbf{Default Value(s)} \\
        \midrule
        nb-nodes           & $100$        \\
        total-ticks        & $100$        \\
        network            & Erdős-Rényi\\
        k-value            & 8          \\
        \( p_o \)          & 0          \\
        \( p_n \)          & from $0$ to $1$ 
        \\
        initial-opinion-metric-value & $0.5$ \\
        opinion-metric-step & 0.10      \\
        activation Threshold (\( \Theta \)) & $0.270$, $0.342$, $0.414$ \\
        echo-chamber-fraction & $0.20$    \\
        \bottomrule
    \end{tabular}
    \caption{Default values for simulation parameters}
    \label{tab:simulation_parameters}
\end{table}

The experiments have been executed using NetLogo 6.2.0 and Python 3.10 on an Intel Core i9-11900KF machine equipped with 32GB DDR4 RAM. Each run of simulation (with Super-Agent) took approximately 3 hours. To obtain significant results we run every simulation 300 times.
The default ABM setup used for both experiments' typology is shown in the table \ref{tab:simulation_parameters}. The used values are the best setup for the conducted experiments. 

\subsection{Results without Super-Agent}
\label{ssec:without}

\begin{figure}[h!t]
    \centering
    \includegraphics[width=8cm]{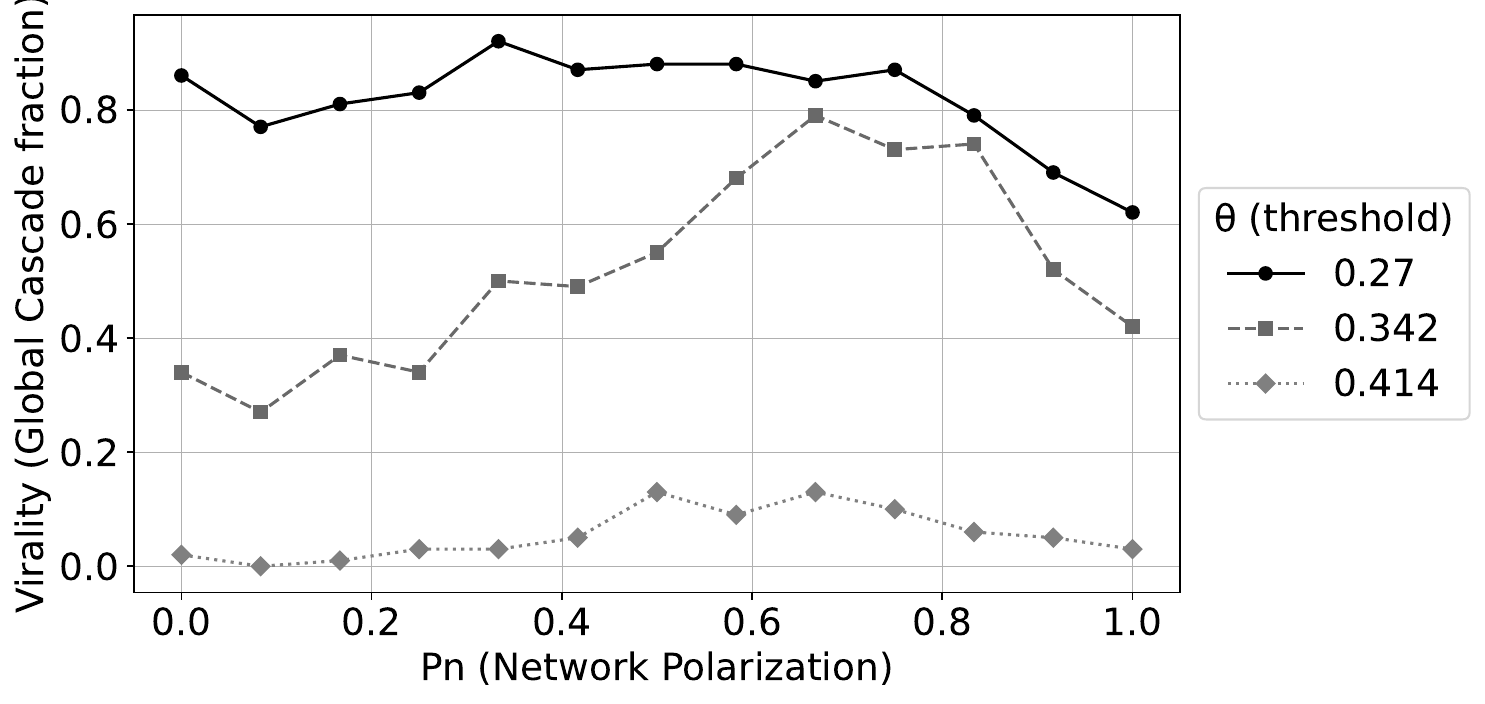}
    \caption{Average Virality Trend without Super-Agent.  $P_O=0$}
    \label{fig:virality_no_sa}
\end{figure}

 For computational and time reasons, we have decided to use the mainly three representative values of $\theta$ (0.270, 0.342, 0.414) considering the values used in \cite{tornberg2018echo}.   
As can be seen in Figure~\ref{fig:virality_no_sa}, when $\theta$ is equal to $0.27$, V remains high (values greater than $0.8$) for $P_N$ values between $0.2$ and $0.8$ and decrease only when the network is highly polarized ($P_n > 0.8$) because, despite the high credulity of the agents, the echo chamber is highly isolated from the rest of the network and has low chance to influence the external agents. 
When $theta$ increases and is equal to $0.342$ V takes on its highest values (values greater than $0.4$) with $P_N$ between $0.6$ and $0.8$ and, as in the previous case, decreases for high network polarization values. Finally, for $P_N$ equal to $0.414$, V appears to be quite linear and little affected by the presence of the echo chamber, regardless of the polarization value of the network, with high values minor than $0.2$ for $P_N$ value between $0.5$ and $0.7$. 

\begin{figure}[h!t]
    \centering
    \includegraphics[width=8cm]{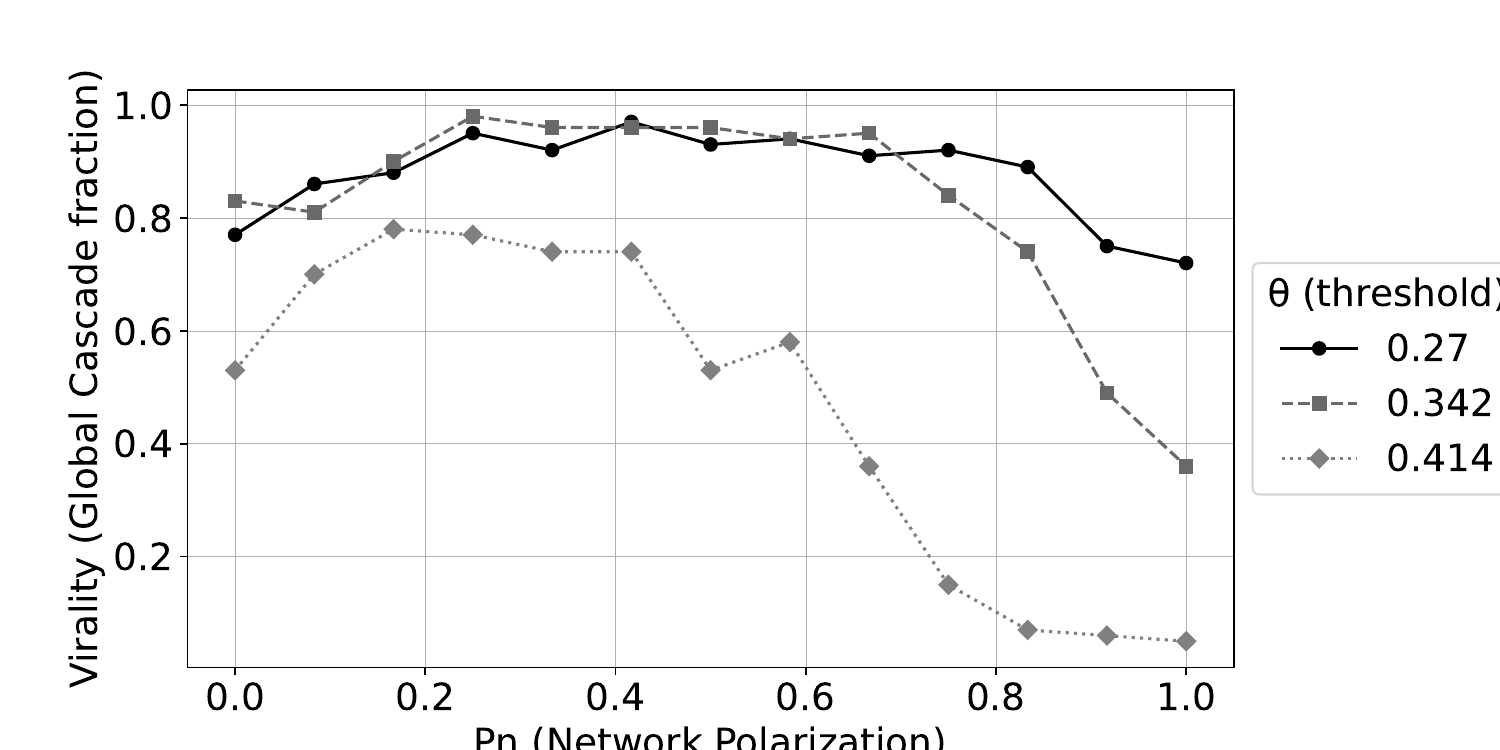}
    \caption{Average Virality Trend without Super-Agent. $P_O=0.27$}
    \label{fig:virality_no_sa_PO_027}
\end{figure}

 Thus, high Virality is achieved when agents are very gullible and the echo chambers have a fair number of connections with the rest of the networks. Especially in such situations, external supervision may be necessary to limit the spread of misinformation. Before turning to the Super-Agent experiments, let us examine in Figure~\ref{fig:virality_no_sa_PO_027} the results obtained considering $P_O = 0.27$. As explained earlier, for the echo chamber the threshold value became $theta - P_O$, increasing the credulity of the agents involved. 
Comparing the results shown in Figure~\ref{fig:virality_no_sa_PO_027} with results with $P_O = 0$ (Figure~\ref{fig:virality_no_sa}) we can see that with $P_O = 0.27$ (as in~\cite{tornberg2018echo}) the overall value of the V is higher for all the $P_N$ and $\theta$ values, confirming how the credulity of the echo chamber has a significant impact on the spread of fake news and how important is external supervision. 

\subsection{Results with Super-Agent}
\label{ssec:with}
In addition to the default parameters reported in table \ref{tab:simulation_parameters}, the parameters needed for the super-agent are listed in Table~\ref{tab:default_values_super_agent}.

\begin{table}[H]
    \centering
    \begin{tabular}{lr}
        \toprule
        \textbf{Attribute} & \textbf{Default Value(s)} \\
        \midrule
        node-range & 0.10 \\
        node-range-static-b & 0.05 \\
        global-warning & True \\
        choose-method & degree \\
        warning-impact & 0.10 \\
        sa-delay & 5,4,2 \\
        \bottomrule
    \end{tabular}
    \caption{Default values for the simulation experiments with Super-Agent}
    \label{tab:default_values_super_agent}
\end{table}

\begin{figure}[h!t]
    \centering
    \includegraphics[width=8cm]{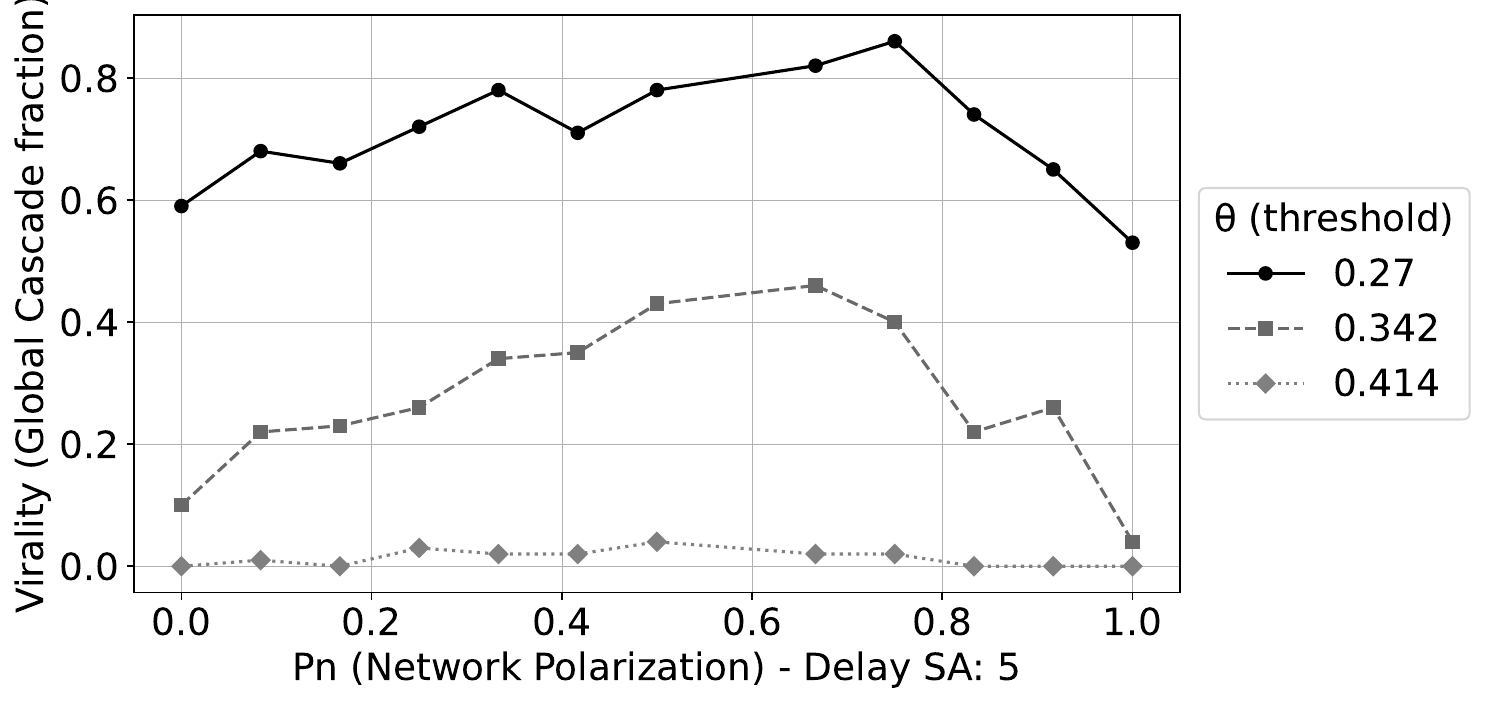}
    \caption{Average Virality Trend with Super-Agent. Action every 5 ticks. $P_O=0$}
    \label{fig:virality_sa_5}
\end{figure}

\begin{figure}[h!t]
    \centering
    \includegraphics[width=8cm]{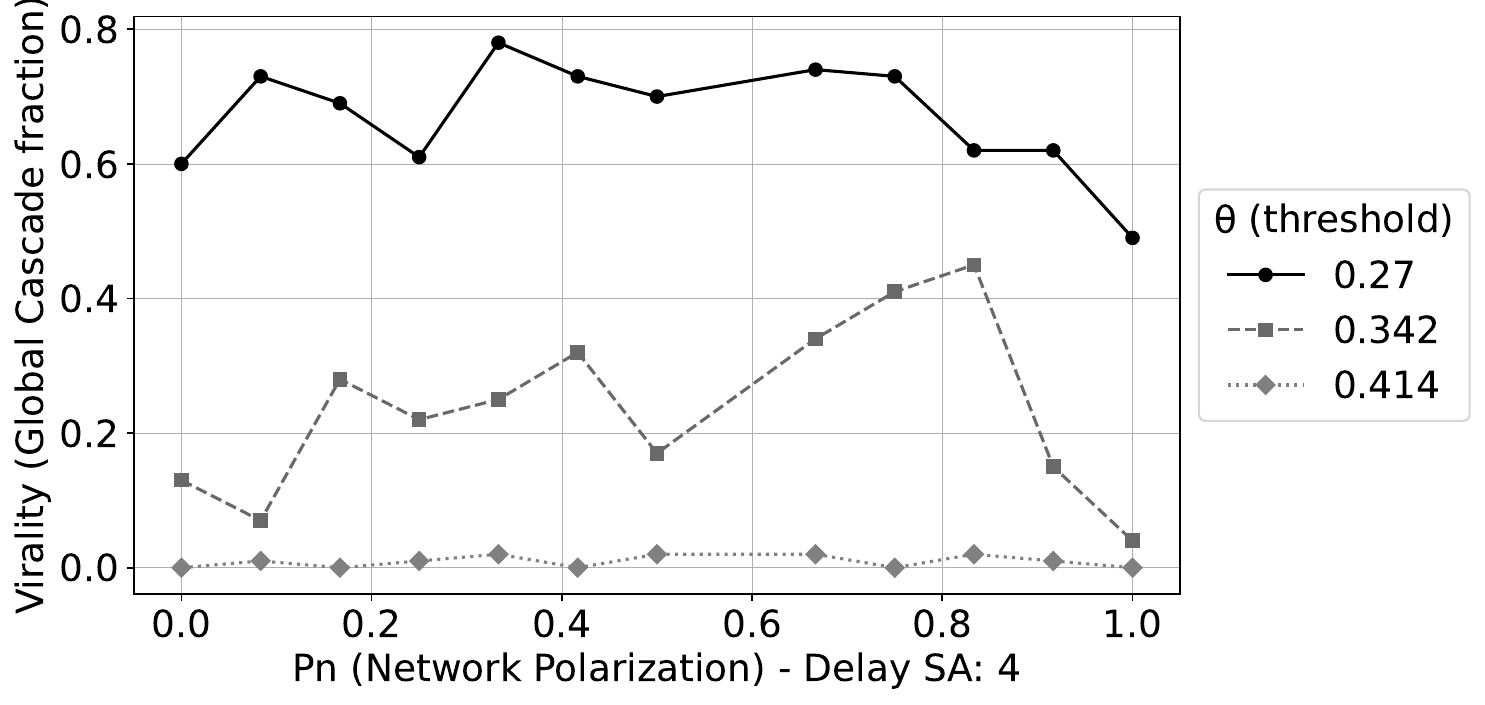}
    \caption{Average Virality Trend with Super-Agent. Action every 4 ticks. $P_O=0$}
    \label{fig:virality_sa_4}
\end{figure}

\begin{figure}[h!t]
    \centering
    \includegraphics[width=8cm]{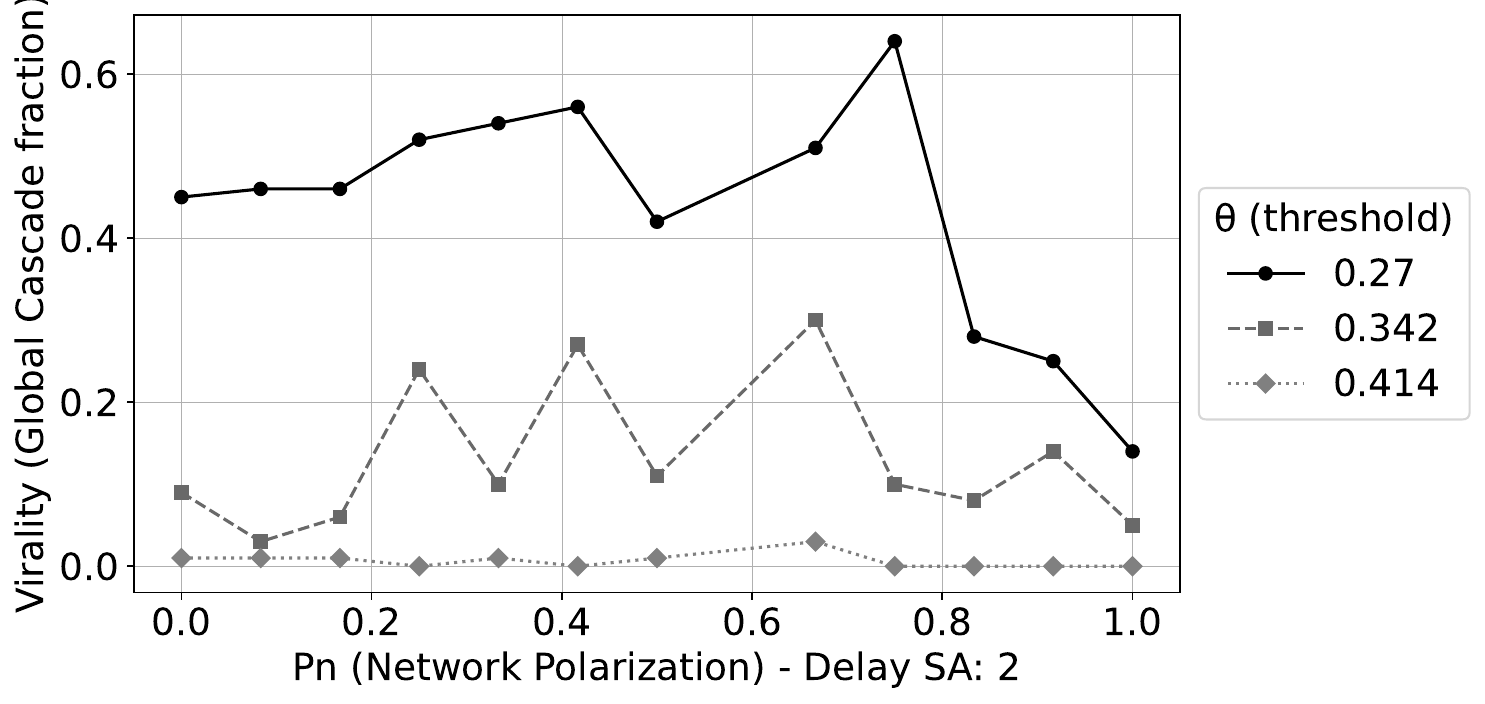}
    \caption{Average Virality Trend with Super-Agent. Action every 2 ticks. $P_O=0$}
    \label{fig:virality_sa_2}
\end{figure}

\begin{figure}[h!t]
    \centering
    \includegraphics[width=8cm]{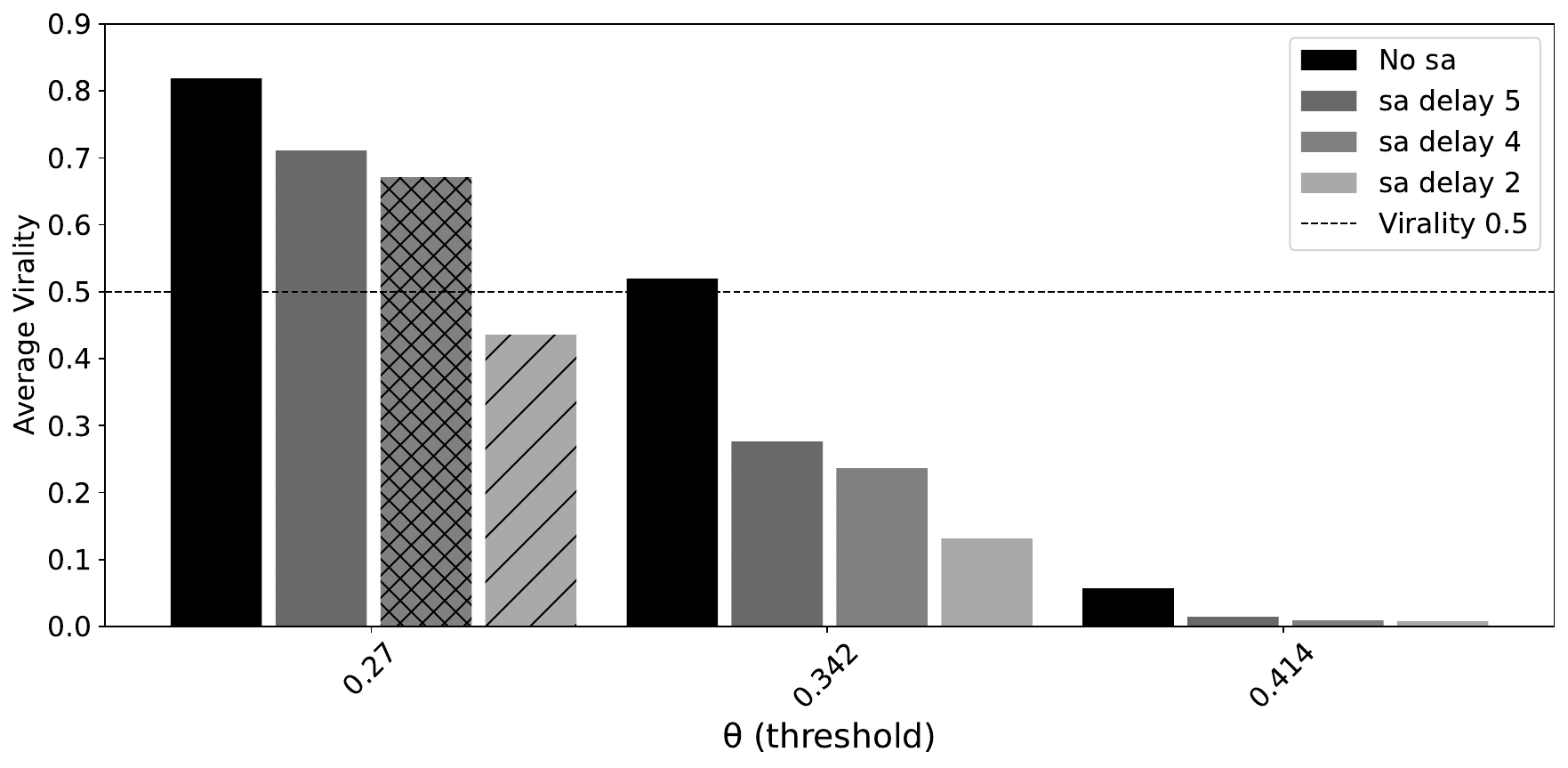}
    \caption{Average Virality Trend with Super-Agent. Differences.}
    \label{fig:virality_sa_differences}
\end{figure}

For this experiment, we ran the simulations three times by changing the {\tt sa-delay} value to understand how the frequency of Super-Agent interventions on the fake news spreading affects the Virality value. 
As we can see in Figures~\ref{fig:virality_sa_5},~\ref{fig:virality_sa_4} and~\ref{fig:virality_sa_2}, lower V values are obtained when the Super-Agent executes more actions in a short time, with a delay of two ticks. In Figure~\ref{fig:virality_sa_5} the trend obtained with $\theta = 0.27$ is very similar to the trend obtained without Super-Agent (see Figure~\ref{fig:virality_no_sa}). Instead, for $\theta = 0.342$, the function takes lower values on average with $P_N$ in the range from $0.4$ to $0.8$, compared with the same experiment without the Super-Agent actions. As shown in Figure~\ref{fig:virality_sa_4}, the same considerations can be made when the {\tt sa-delay} is equal to $4$ and $\theta$ is equal to $0.27$ and $0.342$.
For $\theta = 0.414$, in all the experiments with Super-Agent, V stays low regardless of $P_N$ value, and similar to the trend shown in Figure~\ref{fig:virality_no_sa}. 
In general terms, we can say that the Super-Agent intervention is highly required when the $P_N$ is in the range $[0.4, 0.8]$, where the results are relevant with {\tt sa-delay} set to two. Figure~\ref{fig:virality_sa_differences} compares the average V value between the experiments executed and described above. 
For $\theta = 0.27$ to obtain an average V value lower than $0.5$ is necessary to apply Super-Agent's actions every two ticks. For $\theta = 0.342$, lowering the V value is easier for the Super-Agent, with actions also every five ticks.  Finally, for $\theta$ equal to 0.414, the Super-Agent supervision is not necessary. 
\begin{figure}[h!t]
    \centering
    \includegraphics[width=8cm]{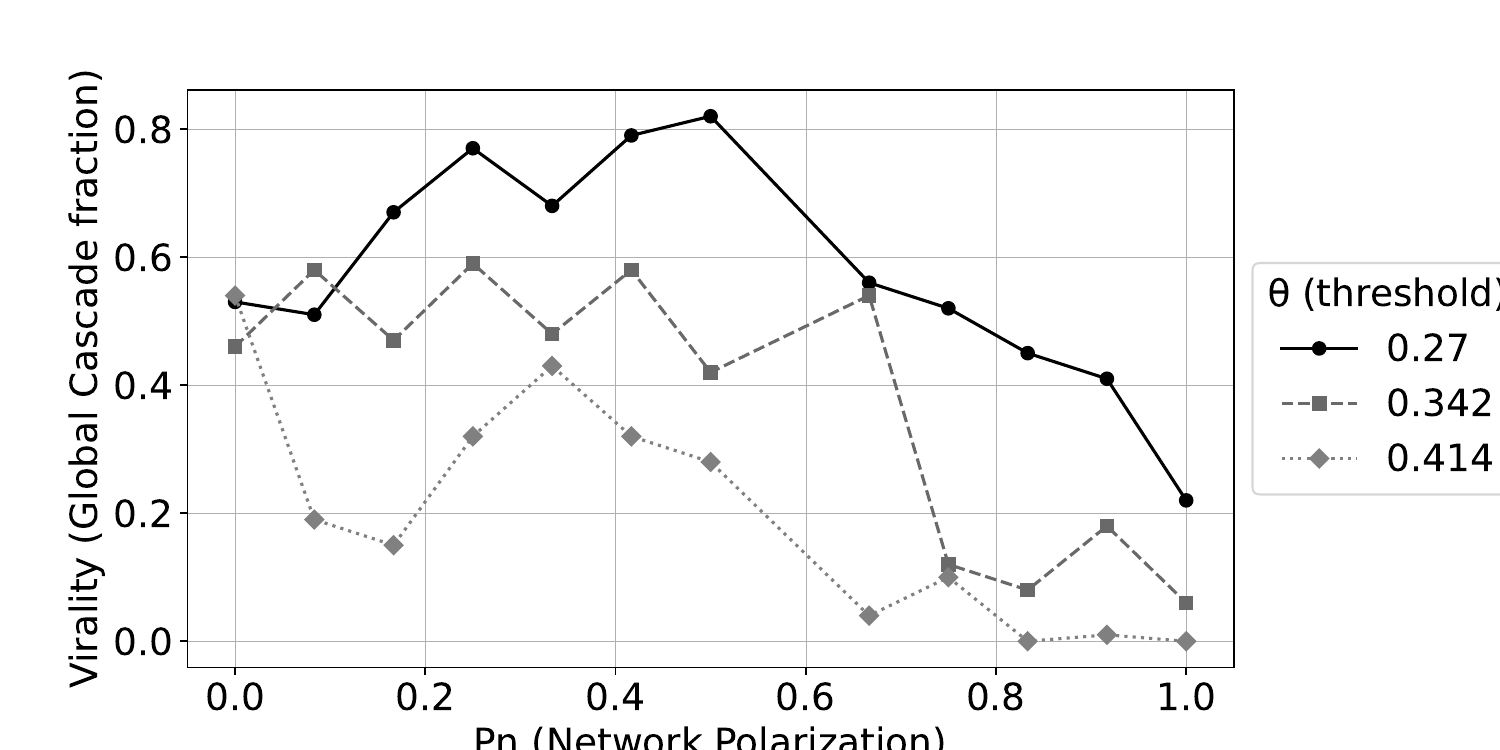}
    \caption{Average Virality Trend with Super-Agent. $P_O=0.27$ and sa-delay=2}
    \label{fig:virality_sa_PO_027}
\end{figure}

Figure~\ref{fig:virality_sa_PO_027} shows the simulation results obtained using $P_O = 0.27$ as done in Figure~\ref{fig:virality_no_sa_PO_027}. In this case, we have used $sa-delay = 2$ because has a major impact on the Virality. The results obtained, compared with Figure~\ref{fig:virality_sa_2} demonstrate how the opinion polarization also plays a crucial role with Super-Agent supervision and how his actions return a good response on the Virality considering the results in Figure~\ref{fig:virality_no_sa_PO_027}.

\section{Conclusion}
\label{sec:conclusion}
The research and experiments presented in this paper show a novel approach to studying and addressing information disorder, particularly fake news, on social networks, marked by the integration of model-driven and data-driven approaches to the analysis of complex social phenomena.
The results of our work unfold on different levels. A first achievement lies basically on the methodological level: the hybrid nature of the research strategy adopted seems to be a promising way to better delve into the complexity of social phenomena policy and rule makers have to deal with \cite{lettieri2016computational}. 
 
On a more substantive level, key findings reveal that the presence of a Super-Agent, operating on DRL principles, significantly impacts the mitigation of fake news propagation. Experiments conducted with various network parameters and intervention frequencies demonstrate a marked reduction in the Virality of fake news, especially in networks with moderate levels of echo chamber effects. This effect is amplified as the intervention frequency of the Super-Agent increases. 
Furthermore, the research underlines the importance of addressing both the content and the context of news in social networks. The study’s success in exploring strategies allowing to reduce misinformation spread highlights the potential of interdisciplinary approaches in addressing complex societal challenges. Future works will head in more directions. An ad hoc analysis be conducted to study the impact that any single action performed by the Super-Agent has on Virality focusing also on dynamical aspects like the impact of timeliness of action on ID evolution or the sensitivity to initial conditions. Moreover, we will focus on the definition of the most effective Super-Agent DRL architecture comparing a set of state-of-the-art DRL models and customized ones.

We will finally focus also on technical issues like the enhancement of the tools implementing the two tiers. A crucial challenge will be that of improving the performances of the ABM environment since simulations quickly become highly demanding from a computational point of view as soon as even minimal changes are made to the complexity of the model (number of agents, interaction structure, complexity of the decision mechanisms implemented in each agent).

\bibliographystyle{named}
\bibliography{ijcai23}

\end{document}